\newif\ifpdf
\ifx\pdfoutput\undefined
  \pdffalse              
\else
  \pdfoutput=1           
  \pdftrue
\fi
\documentclass[aps,prb,twocolumn,superscriptaddress]{revtex4}
\ifpdf
  \usepackage[pdftex,
        colorlinks=true,
        pdftitle={A New face on old code},
        pdfauthor={P.F. Peterson and Th. Proffen},
        pdfsubject={},
        pdfkeywords={NOBUGS2002/024},
        pdfpagemode=None,
        bookmarksopen=true]{hyperref}
  \usepackage{thumbpdf}
\fi
\usepackage{graphicx}
\usepackage{amssymb}

\begin{document}


\title{GUI Tools for an Enhanced User Experience}


\author{P.A. Kienzle}
\affiliation{National Institute of Standards and Technology, Gaithersburg, MD \ 20899, USA}

\date{October 31, 2002}
\begin{abstract}
For instruments with many occasional users, it is important to have
easy to use software.  To support the frequent users it is
important to be flexible.  Using a scripting language to design a
GUI and exposing it to the user allows us to do both.  We present
our work on a GUI for reflectometry data reduction and analysis
written in Tcl/Tk and Octave, with underlying C code for the numerically
intensive portions.  As well as being easier to train new users,
the new software allows existing users to do in minutes what used
to take hours.
\end{abstract}
\maketitle

\section{Introduction}

Reflectivity data reduction and analysis at the NCNR has been a 
mix of various command line tools written in C and Fortran, often 
hidden behind scripts.  Over the past year we have been reimplementing 
the functionality of these tools using a combination of 
Tcl/Tk~\cite{tcl:web} and Octave~\cite{octave:web}.

As we describe reflectometry data reduction and analysis there are
a number of themes to keep in mind.  Software should
do the right thing most of the time.  Software should allow you to
do the wrong thing if necessary.  Software should not be limited to
just those features that are programmed in by hand.  Users should be
free to perform special calclations on their data without losing all
the benefits of the GUI environment.

\section{Reflectometry}


A reflectometer 
consists of an incident beam, slits to control beam
width and beam divergence, attenuators to control flux, the sample
mount, and one or more detectors.  Users can control the slits, the
attenuators, the incident angle of the beam and the reflected angle
to the detector.  The slits are chosen such that the sample is fully
illuminated throughout the entire range of angles.  At very low
angles the slits are fixed slightly open otherwise the beam intensity 
would go to zero.

After reducing the reflectometry data, the user is left with a
reflectivity curve giving the proportion of reflected beam as a
function of Q in reciprocal space.  
In practice, users want to be able to see both Q and angle theta 
throughout the reduction process.

There are four scans that are used to compute the final reflectivity,
as determined by the incident and reflected angles.
Keeping the reflected angle equal to the incident angle
measures the specular reflection.  Keeping the reflected angle slightly
above or below the incident angle measures the off-specular
reflection.  Under the assumption of a perfectly flat sample, the
off-specular reflection is considered to be background noise, and is
to be subtracted from the specular reflection when producing the final
reflectivity curve.  The positive offset and a negative offset
background scans are averaged before subtracting them from the
specular scan. Keeping the incident and reflected angles at zero but
gradually opening the slits measures the slit scan.  This records
the power of the incident beam which you need in order to
normalize the final reflectivity curve.

There is also the rocking curve. Here the incident angle is fixed and
the reflected angle is varied.  If the peak does not occur when the
reflected angle matches the incident angle, then the user knows that
the sample is not mounted properly.  The rocking curve will also
give an indication of the validity of the assumption that the sample
surface is perfectly flat.  Using a set of detectors at different
reflected angles rather than a single one, users can measure an
entire rocking curve at a time.  At present this is used to estimate
the specular and background signals simultaneously so it makes more
efficient use of the instrument.  A topic of research is how to
extract information about the surface of samples which are not
uniform from the grid of rocking curves taken at various incident
angles.

There are a variety of other measurements that are made,
such as fixing slits and angles but varying temperature or field (or
occasionally on NG7, the height of the sample).  Our software can
display these curves, but they do not enter into further reduction
or analysis.

\section{Data selection}

Many raw data files are required to produce one reflectivity curve.
Individual runs are used to cover different parts of the Q range
due to things such as differing counting times, sampling densities
or beam attenuation.  With polarized beam, the data for each
polarization state (A, B, C or D) is taken separately.  In an extreme
case (polarized data with positive and negative Q ranges each split
into several runs) over 100 files may be required to produce a
single reflectivity curve.

The attenuators are placed in the beam by hand, so each time an
attenuator is changed, there will be another file created.  Because
the detector needs time to recover between events, we need attenuators
whenever the rate of neutrons entering the detector is too high.  This
is mainly an issue for slit scans because then the detector is exposed
directly to the beam.

The motor control program only allows motors to be moved by fixed
increments during a single run.  At low angles the user has fixed
slits, so low angle data must be in a separate run.  Depending on 
what they are measuring, users will want to sample some parts of 
the reflectivity curve more densely than others.  Each change in
sampling density requires a new run.  

The measurable reflectivity signal can change by seven orders of
magnitude from below the critical angle where it is one, to high
angles where it is indistinguishable from background.  To get
statistically significant counts throughout the entire range,
different sections are measured for different times.  Whenever the
measurement time changes, a new run is needed.

One instrument has flippers to change the polarization state of
the neutrons.  A complete set of data consists of A,B,C,D files
depending on which of the pair of flippers is activated. Sometimes
reflectivity is measured through front and back surfaces, so each
positive Q run will have a corresponding negative Q run.

For a variety of reasons the same Q range is often measured
several times.  Sometimes it is because the sample is dynamic.
Users will reject the first few passes because the specular
curve is still changing, but they will want to combine the
remaining passes in the final reflectivity.  Plus there are the
usual problems that crop up during an experiment which cause
some runs to be aborted or some ranges to be remeasured.


To make sorting through the data easier, our software automatically
categorizes each file according to what it is trying to 
measure. 
This is easy for reflectometry because that information is completely
determined by the motor movements as recorded in the data file.
The data range is displayed along with the run number.  This makes
it easy to see which runs compose the entire Q-range of the curve
without having to select the files or read from the log book.
Double-clicking the first file in the Q-range automatically selects
all files of the same type which extend the Q-range.  Data taken
with different flippers or different slits or at different
temperature or different field are skipped.  Users can add or
remove files individually, and with a little extra effort they can 
force otherwise incompatible files to be selected together.  An ongoing
theme is to provide convenience without sacrificing flexibility.

Even as a tool for sorting data files without performing any
reduction, experienced users have found our software to be worthwhile.
Being special purpose software it knows how to plot reflectometry
data and automatically normalize for things like monitor count.
What you can do with a double click would take several minutes to
do with command line tools.  The result is that data reduction 
which used to take hours can now be done in minutes.

Even better, our software encourages users to examine the data at each
step of the reduction process.  In one case a subtle problem with the
instrument controller lead to a small discrepency in one of the scans.
Because the data is visible at every step of the way the discrepency
was easy to spot.  With command line and batch files, there isn't a
strong inclination to view the data every step of the way, and the
discrepency wasn't noticed.

For novice users the software is a boon.  Yes they benefit as much from 
the data browsing capabilities as the experienced users, but they
also benefit from the consistency checks which restrict the data that
can be selected together.  Furthermore certain questionable data points
such as those in which the data rate exceeds the known recovery time
for the detector are automatically tagged for exclusion.  While it may
not be obvious to the user why the points are being excluded, it should
be enough of a clue that they will ask a more experienced user what is
going on.  Better that than to quietly accept questionable data.  Users
can override the exclusion easily enough in keeping with the theme of
convenience and correctness without sacrificing flexibility.

\section{Data Reduction}


Once the files are selected the data reduction process is fairly
straight forward.  A set of specular, background and slit scans
are selected. Specular runs are averaged and the average
background is subtracted. The result is divided by the slit
scan and by the incident medium transmission coefficient if the
beam is attenuated by the sample environment.  If the data was
taken with fixed slits at low Q, users need to apply a footprint 
correction to account for the fact that some of the beam spills
over the edges of the sample.

There are of course complications.  For example, the slit scan is
based on slit configuration rather than angle so specular and
background data need to carry slit information along with them so they
can be normalized later on.  That means the data saved in intermediate
files must also record the slits associated with each data point.
This complicates saving and reloading data files.  There are also
the same sorts of complications which arise with data selection:
the software tries to ensure that the scans selected for reduction
are consistent, allowing the user to override if necessary.

Again the GUI interface allows a novice user to easily learn the
necessary steps for data reduction.  The software can keep track
of the state of the data reduction and warn if for example the user
tries to do a footprint correction before selecting the slit scan
which will normalize the data [this is work in progress].

\section{Data Analysis}


After the data has been reduced to a reflectivity curve, the next step
is to try to find a density profile which gives rise to that profile.
If you can change a property of the the sample such as the fronting
medium you can solve the inverse problem directly~\cite{inverse}.  Without
additional constraints however, finding the density profile is an
ill-posed problem since many different profiles can give rise to the
same reflectivity curve.  This situation is exacerbated by a search
space with many, many local minima and an expensive cost function
(about one second per function evalation for a profile of moderate
complexity on my slow machine).

Some fitting tools are model independent in that they try to find a
density profile which generates the reflectivity curve without making
any assumptions about the shape of the profile other than an initial
guess~\cite{splines}.  Other fitting tools are model dependent in that they
assume the sample is made up of particular layers of particular depths
with particular diffusion across the layer boundary.  The user then
codes constrains among the parameters the software finds the best fit.

An example of the latter program is mlayer~\cite{mlayer}.  This is the first
analysis program for which we have provided a GUI interface.  Unlike
other model-based interfaces, ours lets the user directly manipulate
the density profile.  As they drag interfaces, roughnesses and
scattering length densities, users are treated to automatic updates 
of the theoretical density curve overlaid on the reflectivity data 
which they are trying to fit.  Users can also enter specific known 
values into a table of layers.

\section{Tools}

The GUI interface we have been working on over the past year has been
developed primarily in Tcl/Tk.  Tcl is an excellent scripting 
environment in this case because it is simple to learn but still 
powerful.  Because it is a popular language, there are a number of
tools available for it.  Usually, if you need some kind of interface
widget it is a matter of finding one that somebody else has written
rather than writing a new one for yourself.

One big piece missing from Tcl/Tk for scientific programming is that
it does not deal well with vectors and matrices.  There are some
pure Tcl implementations of matrix operations but they are too slow.
The BLT package provides vectors and some vector operations, but
many operations are not available.  Instead we use the Octave
numerical environment as a compute engine.  Again it is a simple
language to learn, and again it is similar enough to Matlab that are
large number of tools are available for it.  Usually, if you need to
solve some kind of numerical problem it is a matter of finding a
solution that someone else has written.

While not a full featured publication quality graph layout
application, the graph widget we use (BLT) provides enough control to
make an excellent data browser.  Being tightly bound to the rest of
our GUI allows us to implement certain conveniences which we could not
easily do with a separate graphing application such as switching from
log to linear by clicking on the axis or displaying the point under
the cursor as both Q and angle coordinates which are very useful in
our application.

Because we are using a scripting language, we can open up a console
which allows us to enter commands directly in that language, including
commands for manipulating data that is shown on the screen.  For
example, certain specialized processing of the counts may be required
to correct for an instrument error.  These procedures only need to be
applied to a few files (e.g., 3 months worth the runs) so it isn't reasonable
to expect support for the specialized processing to go into the general
reflectivity reduction software.  Instead they should be able to modify
the affected files by hand (e.g., using a correction script) and continue
processing the data using the usual gui interface.

Once the experiment is running smoothly, data is taken very regularly.
In that case it should not be necessary to select all the data by hand
for reduction, but instead be able to write a script which reduces all
the data, showing a few key steps along the way so that the user can
be sure that the data is reasonable.  If it is not, the user should be
able to take corrective action using the usual user interface then
continue with the script. The script should also allow the user to
perform steps that cannot be automated, such as selecting the linear
part of the fixed slit region used to fit the footprint correction.
In this way, processing data sets can be reduced from minutes to
seconds.  Even more importantly, instrument responsibles can set up
automated procedures for their users to follow for particular kinds of
experiments.  If all goes well, they should then be able to process
their data with little hassle or detailed knowledge of the usual
sequence of steps required.  

Providing a scriptable interface is a medium term goal for the
project.  We have not yet formalized the underlying data manipulation
in a way that makes this convenient. An issue with all GUI programming, 
no matter what the language is that it is by nature anathema to 
modular programming.  Because the user is free to wander backward 
and forward through the process, it is difficult to implement the 
procedural aspects of the interface.  For example, footprint 
correction requires that the user fit a curve to the flat portion 
of the onset of the background subtracted, slit scan divided data.  
However, changing the background runs may change the position of 
the flat portion of the onset, so the footprint that the user specified 
is no longer applicable.  So somehow selecting new background data 
has to signal that the footprint correction used elsewhere in the 
program is now invalid and should be ignored.  This two way flow of
information is very disruptive.

In my experience Tcl/Tk facilitates hiding the interconnections better 
than most.  Individual widgets can take a variable name as a parameter
rather than the current variable value.  They then put a trace on
the variable so that the widget is notified when the value is read,
written or deleted.  Using techniques like this, we should be able
to make the GUI independent of the scripting language, so that new 
gui elements can be added without updating the existing gui elements
or the scripting interface.

Again, because it is a scripting environment it is relatively easy
to support extensions to the environment.  As the project has grown
from supporting one data format (NG-1) to four (NG-1, NG-7, X-RAY,
NG-7 with position sensitive detector), it has become necessary to
modularize the file loading process.  Adding support for a new file
type now consists of writing two functions, one to quickly load
the header so that the file can be categorized, the data range
displayed, and constraints applied to the selection process, and 
another to load the data and do initial conversions to Q vs intensity.
There is also some code to associate file extensions with the file
categorization function which needs to be changed in several places,
so further refactoring is necessary.  Also, as different instruments
have different capabilities and limitations, some aspects of the
reduction process will need to change elsewhere in the program.

\section{Conclusions}

While not yet complete, our data reduction and analysis tools lead to
a marked increase in productivity.  So much so that even long time
users with dozens of specialized scripts for data reduction are
happy to convert to using them.  Through careful design we have
managed to provide convenience and flexibility in the same package.

While a concern throughout the development, modern PC's are able to
handle the performance penalty of running scripted applications.  The
user interface is adequate even on a relatively old pentium II, 300Mz
computer.  In the case of mlayers, the GUI is bound by the cost of 
generating the reflectivity curve for the layers so I would not 
anticipate a significantly enhanced user experience if the whole 
interface were translated into C.

Tcl/Tk plus Octave has proven itself to be a fine platform for rapid
development of scientific applications.  Interface ideas can be 
tested quickly and the more promising ones can be developed more fully 
without too much overhead.  This allowed us to experiment with a number 
of different interfaces in less time than we would take to produce a 
single interface in C.

\bibliographystyle{aip}

\end{document}